# Monitoring of neoadjuvant chemotherapy through time domain diffuse optics: Breast tissue composition changes and collagen discriminative potential

NIKHITHA MULE[1,3]*, GIULIA MAFFEIS[1], RINALDO CUBEDDU[1], CAROLINA SANTANGELO[3], GIAMPAOLO BIANCHINI[4,5], PIETRO PANIZZA[3] AND PAOLA TARONI[1,2]

[1]*Politecnico di Milano, Dipartimento di Fisica, Piazza Leonardo Da Vinci 32, 20133, Milano, Italy;*
[2]*Consiglio Nazionale delle Ricerche, Istituto di Fotonica e Nanotecnologie, Piazza Leonardo da Vinci 32, 20133, Milano, Italy;*
[3]*Scientific Institute (IRCCS) Ospedale San Raffaele, Breast Imaging Unit, Via Olgettina 60, 20132, Milano, Italy;*
[4]*Scientific Institute (IRCCS) Ospedale San Raffaele, Department of Medical Oncology, Via Olgettina 60, 20132, Milano, Italy;*
[5]*School of Medicine and Surgery, Università Vita-Salute San Raffaele, Via Olgettina 60, 20132, Milano, Italy;*
*nikhitha.mule@polimi.it

## ABSTRACT

The purpose of this clinical study is to test broad spectral range (635-1060 nm) time domain diffuse optical spectroscopy in monitoring the response of breast cancer patients to neoadjuvant chemotherapy (NAC). The broadband operation allows us to fully analyze tissue composition in terms of hemoglobin, water, lipids and collagen concentration, which has never been systematically studied until now during the course of therapy. Patients are subjected to multiple breast optical imaging sessions, each one performed at different stages of NAC, both on tumor-bearing and contralateral healthy breast. We correlate the optical results with conventional imaging techniques and pathological response. Preliminary outcomes on 10 patients' data show an average significant reduction in the concentrations of oxy-hemoglobin (-53%, $p = 0.0020$), collagen (-36%, $p = 0.0039$) and water (-15%, $p = 0.0195$), and increase in lipids (+39%, $p = 0.0137$) from baseline to the end of therapy in the tumor-bearing breast of patients who responded to therapy at least partially. With respect to scattering, the scattering amplitude, *a*, increases slightly by the end of the therapy compared to the baseline, while the scattering slope, *b*, shows no significant change. Some change in constituents' concentrations was noticed also in the contralateral healthy breast, even though it was significant only for oxy-hemoglobin concentration. Collagen seems to be the only component distinguishing between complete ($p = 0.0423$) and partial responders ($p = 0.1012$) by the end of 2-3 weeks after the start of therapy, suggesting that it could serve as a potential biomarker to measure NAC effectiveness early during treatment. These initial results are consistent with those of standard medical modalities and highlight the sensitivity of the technique to changes that occur in breast composition during NAC.

## 1. Introduction

NeoAdjuvant Chemotherapy (NAC) refers to administering chemotherapy prior to surgical intervention in order to downsize the tumor, increasing likelihood of proceeding with breast conserving surgery, to measure early in-vivo response to the treatment and to examine the role of response to neoadjuvant treatment as a predictive marker [1,2]. It could also lead to improved cosmetic outcomes and reduced postoperative complications, such as lymphedema.

Additionally, it enables the early exploration of novel treatment strategies by utilizing pathological complete response (pCR, i.e. the complete disappearance of the tumor in the breast and lymph nodes) as a surrogate endpoint for assessing event-free and overall survival [3].

However, studies have demonstrated that approximately only 20% (average across various cancer subtypes) of the overall patients receiving the NAC therapy show pCR, while the rest of them show partial or no response [4,5]. The early assessment of cancer response to this therapy is essential as it is correlated to patient's survival rate. Therefore, there is a clinical need to predict the therapy response as early as possible and identify non-responders, in order to avoid subjecting patients to any futile medical treatments and, if necessary, to change the treatment strategy in time.

The state-of-the-art monitoring techniques include imaging modalities like x-ray mammography, Ultrasound (US), Magnetic Resonance Imaging (MRI) and Positron Emission Tomography (PET). While x-ray mammography and US may be utilized, they do not readily provide metabolic and physiological information, which could be the key parameter that reflects initial changes rapidly. Further, x-ray mammography uses ionizing radiation that can have adverse effects when used for periodical application like therapy monitoring [6] and US imaging is not sensitive to mild changes that occur in tumor morphology immediately after therapy initiation [7]. Though MRI and PET can give better information on lesion characterization by providing high-resolution structural images and functional information by detecting metabolic activity respectively, they are very expensive. In addition, MRI comprises a very long examination time while patients also experience discomfort due to the prone position required for this exam and PET uses radioactive tracers for imaging [8–11]. Therefore, an inexpensive, non-invasive, non-ionizing imaging modality sensitive to metabolic and physiological changes in the breast would be of critical importance.

Diffuse Optical Spectroscopy (DOS) is being explored by many research groups to understand its potential in this regard. Being a non-invasive technique and having the ability to provide an estimation of tissue composition and microstructure based on absorption and scattering properties, in the case of breast cancer DOS has shown promising applications in risk assessment [12–14], lesion nature identification [15–22], therapy monitoring and early prediction of therapy outcome [23–29], as we will investigate in this paper. It is important to underline that the output of DOS measurements is richer in terms of physiological informative content than morphological one. In fact, due to the spectral range of use, DOS cannot match the spatial resolution of conventional imaging techniques, remaining limited to about 1 cm. This technique thus fits into the trend of multimodal approaches favored and encouraged in recent times [30–35].

As NAC is a form of pre-surgical therapy, it presents the unique opportunity for *in vivo* imaging techniques to identify specific metrics that can help predict the response to it. In the case of DOS, it allows for the retrieval of, for example, hemoglobin, lipid, water and collagen concentrations to explore the correlation between changes in the breast composition and the impact of the therapy over time. This way, diffuse optical mammography could become a powerful predictive technique for the assessment of the effectiveness of NAC therapy.

With this purpose, we are currently carrying out a clinical trial, "NADOPTIC - Optical monitoring of neoadjuvant chemotherapy and prediction of pathological complete response in breast cancer patients", at San Raffaele Hospital in Milan, Italy. Its principal goals are to monitor the compositional changes in the breast tissue due to therapy at various time points through time domain DOS, correlate the optical results with conventional imaging techniques and pathologic outcome, and ultimately predict the effectiveness of NAC in breast cancer patients. This is to discriminate responders from non-responders as early as possible after the therapy initiation through quantitative non-invasive assessment of relevant biomarkers.

Many research groups have already demonstrated the potential of this technique for this application in continuous wave and frequency domain optical techniques [23,36]. In time domain, studies were limited to wavelengths lower than 900 nm [37]. We claim that, to the best of our knowledge, this is the first clinical trial that is being conducted using a time-resolved optical mammograph for the application of neoadjuvant chemotherapy monitoring and outcome prediction in the broad spectral range from 635–1060 nm. The time domain approach up to 1060 nm allows for the investigation of collagen in addition to hemoglobin, water and lipids concentrations, with a better signal-to-noise ratio and limited absorption-scattering cross-correlation [38]. The behavior of collagen has never been systematically studied so far in this type of study and could be of importance given its role in the onset and progression of cancer itself [39].

## 2. Materials and Methods

### 2.1 Instrument Set-Up

Fig. 1(a) shows the schematic design of the optical mammograph that was developed at Politecnico di Milano and being used in this study [40], [41]. The entire set-up is a stand-alone instrument (50 cm W x 80 cm D x 140 cm H), mounted on wheels as shown in Fig. 1(b) and is approved by the Italian Ministry of Health for use in a clinical environment. It works under time domain DOS technique, operates in transmission geometry and collects projection images of the compressed breast, in a scheme similar to that of conventional x-ray mammography. It consists of seven pulsed picosecond lasers as light sources emitting at 635, 680, 785, 905, 933, 975 and 1060 nm (LDH-P, PicoQuant GmbH, Germany), selected based on characteristic features of the breast chromophores' spectra: oxy and deoxy hemoglobin, water, lipids, and collagen. The breast is raster-scanned by the light transmitted through an injection fibre, which is axially aligned with a detection probe composed of 8 SiPMs (Silicon PhotoMultipliers, S13360- 1350PE, Hamamatsu Photonics, Japan). The output signals are then amplified and directed to a Time-To-Digital Converter (MultiHarp 150 8N from PicoQuant GmbH, Germany) that constructs the distributions of photons' times of flight inside the tissue.

The whole breast is scanned with measurements performed after every 2 mm along both axes, with an acquisition time for each pixel equal to 25 ms. The overall scan time depends on the area of the compressed breast that needs to be imaged, but it is typically around 5 min. The compression unit can be rotated by an angle up to 90° in both clockwise and counter-clockwise direction, enabling the imaging of both breasts in the cranio-caudal (CC) as well as medio-lateral or oblique (OB, at 45°) views. Further details about the instrument can be found in references [40] and [41].

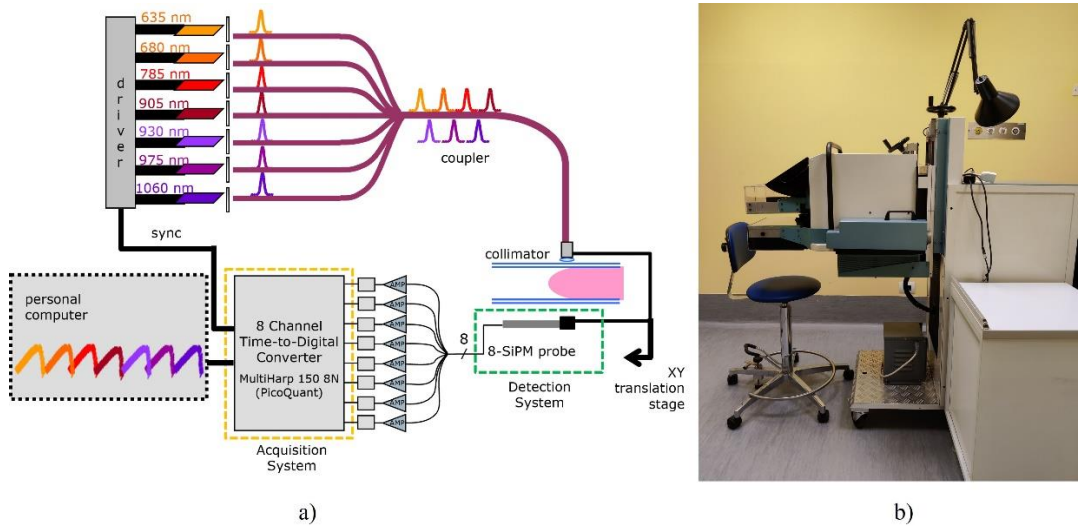

a)                                                                                   b)

Fig. 1. Seven-wavelength time-resolved diffuse optical mammograph: a) Schematic design, b) Picture.

## 2.2 Clinical Trial

### 2.2.1 Clinical Protocol Registration

This clinical study is being conducted at San Raffaele Hospital, Milan, Italy with the approval from its ethical committee and in accordance with the principles laid down by the 18[th] World Medical Assembly (Helsinki, 1964) and the ICH guidelines for Good Clinical Practice (ClinicalTrials.gov ID: NCT04767659). Subjects are enrolled into the study only after they have accepted and signed the informed consent form.

### 2.2.2 Patients and Study Design

Subjects participating in the study are selected based on certain inclusion and exclusion criteria. The inclusion criteria comprise of the signature of the informed consent, being over 18 years of age, a diagnosis of breast cancer prescribed to NAC and a tumor size greater than 1 cm (to comply with the limited spatial resolution of the DOS technique, at least for this initial study). The exclusion criteria include pregnancy of the subject or breastfeeding, previous surgery of the same breast, presence of distant metastasis and presence of any breast implants.

Besides the standard radiological monitoring of NAC (that includes x-ray mammography, US, PET and MRI to keep track of the tumor size and metabolism), the subjects participating in this clinical study undergo 6 measurements with our diffuse optical mammograph: i) a baseline before starting NAC, ii) 2-5 days post NAC start, iii) 6-8 days post NAC start, iv) 2 weeks post NAC start, v) at mid-treatment and vi) at the end of the treatment, as shown in Fig. 2. In each session, 4 scans are acquired, to probe the combinations of breast (right – R – and left – L) and view (CC and OB), thus collecting information both on the tumor bearing breast and on the contralateral healthy one. In this ongoing clinical trial thus far 15 patients have been enrolled, out of which 10 patients have completed the study as summarized in Table 1. Of the 5 patients who could not complete their study, one was tested positive to COVID-19, another was diagnosed with metastasis after 3 measures, while the rest 3 withdrew from the study.

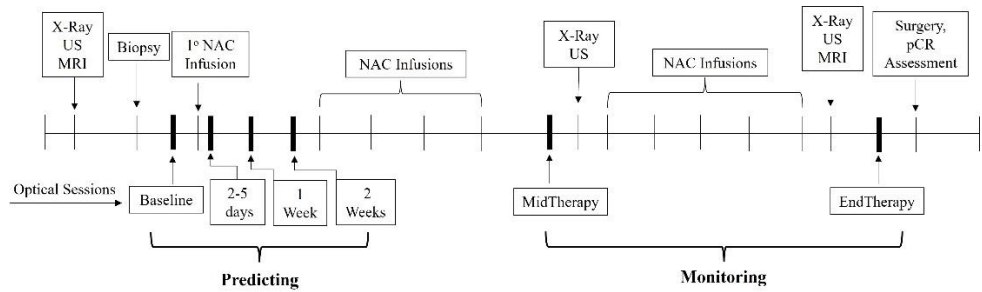

Fig. 2. Timeline of NADOPTIC Clinical Trial, showing medical events and conventional imaging sessions (top) and optical measurements' sessions (bottom).

**Table 1. Summary of tumor characteristics and pathological responses in study cohort.**

| ID | Position | Initial size [mm] | Final size [mm] | Histopathology Result | | | |
|---|---|---|---|---|---|---|---|
| | | | | Response | Residual Cancer Burden Score (RCB) | RCB Class | Grade |
| 2 | R, QQEE | 52 | <1% cellularity | Partial response | 1.159 | I | 2 |
| 3 | L, QSE | 23, 30, 35, 13 | 0 | pCR | - | - | - |
| 4 | R, QSE | 28 | 16 | Partial response | N.A. | N.A. | 3 |
| 5 | L, QQSS | 35 | 0 | pCR | - | - | - |
| 6 | L, QQEE | 70 | 59 | Partial response | 3.679 | III | 2 |
| 7 | R, QQSS | 43 | 0 | pCR | - | - | - |
| 8 | R, retro-areolar | 40 | 16 | Partial response | 2.059 | II | 2 |
| 9 | R, QQSS | 40 | 0 | pCR | - | - | - |
| 11 | R, QQEE | 20, 9, 13, 4 | 11 | Partial response | 1.191 | I | 3 |
| 12 | R, QSE | 18 | 4 | Partial response | 1.25 | I | 2 |

In Table 1, the initial tumor size refers to the longest extension taken from the conventional imaging techniques' reports, predominantly MRI, supplemented by x-ray mammography or US when necessary and the final tumor size is based on the histology of the surgical specimen. When multiple values are reported, it means that a cluster of very close tumors has been detected. From an optical point of view, they will be treated as a single global tumor, as we do not expect DOS to be capable to resolve them.

Considering all patients, the average age is $52 \pm 11$ years old and the average initial tumor size is $39 \pm 16$ mm (for clusters, the biggest nodule has been considered).

In general, the tumor position is defined using a 3/4-digit code, referring to the breast quadrant (Q, first digit) of interest: inferior (I)/superior (S) – second digit - and internal (S)/external (E) – third digit. Double Q means that the tumor is located between the quadrants appointed by the subsequent letters.

All the patients received a biopsy prior to treatment to confirm the cancer diagnosis, which also provided information on type of cancer and tumor receptor status. Various multidrug and/or multicycle regimens were administered to the patients based on that.

After surgery, the resected tumor tissue was sent for histopathological assessment for a detailed report which includes pCR to NAC, residual cancer burden (RCB) index, tumor grade, stage and other relevant findings. The pathology report establishes the ground truth for determining the complete, partial, or non-response to NAC. The histopathology records reported that all these 10 patients have responded to the therapy, wherein 4 patients achieved pCR, while 6 patients had partial response. This could also be noted from the histopathology results reported in Table 1.

### 2.3   Data Analysis

The time-resolved curves were analyzed applying a spectrally-constrained fit, which is a single-step procedure that analyses all the transmittance curves at different wavelengths at once, directly evaluating the breast chromophores' concentrations and the scattering parameters [42]. The analytical solution of the diffusion approximation to the transport equation for an infinite homogeneous slab with extrapolated boundary conditions can be rearranged as product of two terms, each one depending on a single optical parameter as given by [Eq. (1)]:

$$T(t, \mu_a, \mu_s') = S(t, \mu_s',)e^{-\mu_a v t} \quad (1)$$

where $v$ is speed of light in the medium, t is time, $\mu_a$ is absorption coefficient and $\mu_s'$ reduced scattering coefficient. From Beer Law we have,

$$\mu_a(\lambda) = \Sigma_i C_i \varepsilon_i(\lambda) \quad (2)$$

where $\varepsilon_i(\lambda)$ is the extinction coefficient, $C_i$ is the concentration of the i[th] constituent at wavelength λ. Using Mie scattering theory , we have

$$\mu'_s(\lambda) = a\left(\frac{\lambda}{\lambda_0}\right)^{-b} \quad (3)$$

where $\lambda_0$ = 600 nm, $a$ and $b$ are scattering amplitude and slope respectively.

Upon replacing [Eq. (2)] and [Eq. (3)] in [Eq. (1)], we obtain an expression depending on the wavelength that directly relates the tissue constituents' concentrations and the scattering parameters a and b to transmittance $T(t, \mu_a, \mu'_s)$ as shown by [Eq. (4)]:

$$T(t, \lambda; a, b, C_i) = S\left(t, a\left(\frac{\lambda}{\lambda_0}\right)^{-b} e^{-vt \sum_i C_i \varepsilon_i(\lambda)}\right) \quad (4)$$

Therefore, the fit directly leaves as free parameters lipid, water, collagen oxy- and deoxy-hemoglobin ($HbO_2$ and $Hb$, respectively) concentrations and the scattering amplitude, *a*, and slope, *b*. From $Hb$ and $HbO_2$, the total hemoglobin content, $tHb = Hb + HbO_2$ and the oxygen saturation level $SO_2 = HbO_2/tHb$ were calculated. This data acquired from the optical imaging of the breast is visualized in the form of bidimensional maps that represent the spatial variation of each parameter using a custom-made code in MATLAB (Release 2012b, The MathWorks, Inc., Natick, Massachusetts, United States). The concentrations from 4 measures in each session (CC and OB views of both tumor bearing breast and contralateral healthy breast) were retrieved by selecting a rectangular region of interest (ROI) enclosing approximately the tumor area at baseline. The tumor area was known from external references like x-ray mammography, US, and MRI. The size of the ROI was kept the same for all 6 sessions and it encloses at least 25 pixels to cover an area of 1 cm$^2$ (minimum tumor size satisfying the inclusion criteria). The composition of breast tissue retrieved is averaged over all the pixels present in the ROI. The values reported in this paper were obtained by taking the average on both views (CC and OB) to depict an overview of the mean behavior. Similar results were observed when CC and OB views were considered independently. The coefficient of variance for the concentration values retrieved between CC and OB views in each session for each breast was found to be in the range of 2%-13%. Considering that all patients responded at least partially to therapy, keeping the size of the ROI unchanged for different NAC sessions implies that our average values include progressively more and more healthy tissue, and this has to be taken into account when variations are observed over time, as they may come both from a change in composition of the tumor tissue and from a mere shrinking of the tumor volume.

The patients were not distinguished as per cancer subtypes or therapy regimes while taking the average due to their limited number available so far. The Wilcoxon signed-rank test and paired Student *t*-test were employed to determine statistical significance between two groups (for example, tissue concentrations at baseline *vs* end therapy), with $p < 0.05$ considered for significance [43]. The normality test was done using Shapiro-Wilk test before applying the Student *t*-test.

## 3. Results and Discussion

In this Section, we present the compositional alterations observed in breast tissue following NAC, organized into two distinct aspects. Firstly, in Section 3.1, we present variations in the concentrations of various chromophores due to the therapy in the breast tissue at its extreme timepoints: baseline and end of therapy in both tumor bearing breast and contralateral healthy breast. Secondly, in Section 3.2, we provide an analysis of collagen, a component that has not been previously explored in this type of studies, but might hold significant potential for research due to its known involvement in the initiation and progression of cancer [39].

*3.1 NAC effects on breast tissue*

3.1.1 Tumor-bearing breast

We present an overview of the dynamic changes in tumor composition by focusing on the variations in concentrations between baseline and end-therapy time points. This initial analysis offers a broad perspective on the treatment-induced alterations in tumor components, serving as a foundational understanding of the response.

The box plots in Fig. 3 outline the average absolute concentrations of various breast constituents at baseline and end-therapy sessions on 10 patients in the ROI on the tumor bearing breast averaged on CC and OB views. A higher mean total hemoglobin, water and collagen content can be observed at the baseline (before the start of NAC therapy) than at the end-therapy session. The opposite trend is observed in lipids. Both *tHb* and *SO$_2$* decrease markedly with therapy, due to a strong decrease in *HbO$_2$*, while on average *Hb* is not significantly affected. As for scattering, on average, $a$ increases slightly by the end of the therapy compared to baseline, while the scattering slope $b$ shows no significant change.

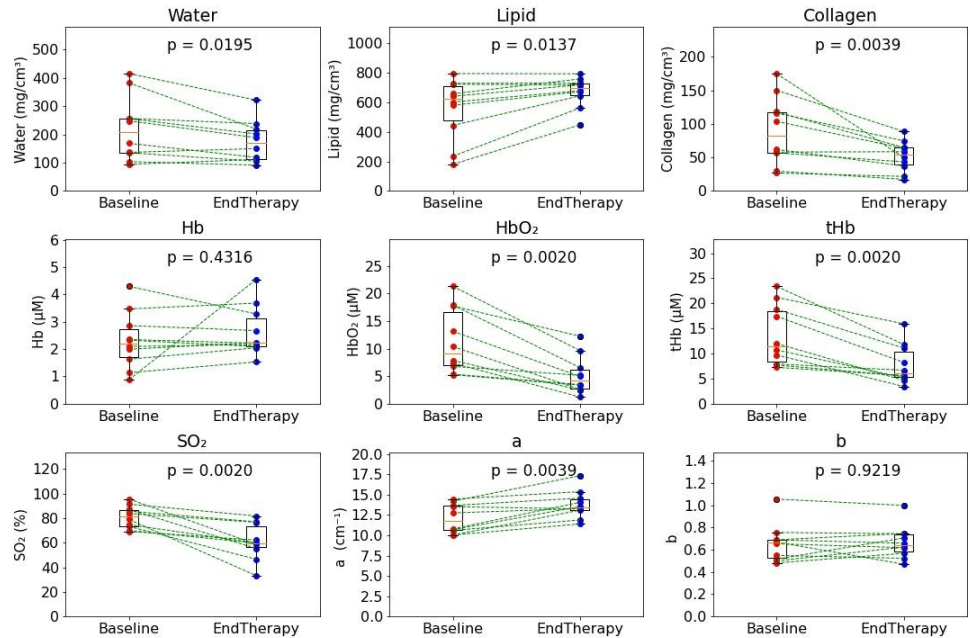

Fig. 3. Tissue constituents' concentrations, blood parameters, and scattering parameters at baseline and end of the therapy for 10 patients in tumor bearing breast averaged over CC and OB views. *p*-values for the Wilcoxon signed-rank test are shown.

More quantitatively, as reported in Table 2, the Wilcoxon signed-rank test on this data showed that at end-therapy the value of all optically derived-parameters is statistically different from that at baseline, indicating that in the tumor region the composition (lipid, water, collagen), vascularization (*tHb* and *SO$_2$*), and microscopic structure (scattering amplitude $a$) change significantly with effective treatment. Only exceptions are deoxy-hemoglobin and the scattering parameter $b$. These exceptions could be due to their small numbers, thus determining significant relative errors also for limited variations in absolute values. Considering the mean of 10 patients and averaging on both CC and OB views, there is an overall significant reduction in the concentrations of total hemoglobin (-41%), water (-15%) and collagen (-36%) and an increase in lipids (+39%) and $a$ parameter (+15%) by the end of therapy in the tumor bearing breast, indicating a response to the treatment, which makes the average tissue condition (composition, vasculature, structure) more similar to what is expected from heathy tissue based on optical measurements [8].

**Table 2. Absolute tissue constituents' concentrations, blood parameters, and scattering parameters in tumor bearing breast averaged across 10 subjects at baseline and end therapy sessions, and over CC and OB views. [%Change = (EndTherapy_Concentration – Baseline_Concentration)/Baseline_Concentration].**

|  | Baseline | EndTherapy | Average change (%) | p-value |
|---|---|---|---|---|
| Water (mg/cm³) | 219 ± 106 | 175 ± 69 | -15% | 0.0195 |
| Lipid (mg/cm³) | 559 ± 198 | 673 ± 97 | 39% | 0.0137 |
| Collagen (mg/cm³) | 90 ± 48 | 52 ± 22 | -36% | 0.0039 |
| Hb (μM) | 2.3 ± 1.0 | 2.7 ± 1.0 | 45% | 0.4316 |
| HbO$_2$ (μM) | 11.3 ± 5.6 | 5.2 ± 3.3 | -53% | 0.0020 |
| tHb (μM) | 13.6 ± 5.7 | 7.8 ± 3.8 | -41% | 0.0020 |
| SO$_2$ (%) | 81.0 ± 9.0 | 61 ± 13.9 | -24% | 0.0020 |
| *a* (cm$^{-1}$) | 12.1 ± 1.7 | 13.7 ± 1.6 | 15% | 0.0039 |
| *b* | 0.66 ± 0.16 | 0.67 ± 0.14 | 4% | 0.9219 |

Besides, these findings correlate to the histopathological reports, that show that all 10 patients responded to the therapy (either partially or completely), and associate to medical literature, that states that blood, water, and collagen concentration is higher in malignant tumors compared to healthy tissue, which may be due to the interplay between angiogenesis, hypermetabolism and intravascular pressure in the tumor [44,45]. On the contrary, the lower lipid content in tumors compared to normal tissues could be due to the replacement of breast adipose tissue by the tumor cells. Therefore, after an effective NAC therapy, the final blood, collagen and water concentration is expected to be lower, while the lipid content is expected to be higher compared to the baseline concentration.

Based on the decrease in collagen and increase in lipid content with therapy, we would expect a flatter reduced scattering spectrum at the end of therapy, reflecting in a higher scattering slope *b* and lower scattering amplitude *a* (which corresponds to the reduced scattering at 600 nm). The fact that instead we observe some increase in *a* and no significant trend in *b* suggest that, apart from changes in composition, some modification in the microscopic tissue structure in caused by NAC [12,46,47].

Overall, it has to be noted that, for this initial investigation, a homogeneous tissue model was used for data analysis, leading to average tissue values of the estimated parameters over the probed volume. Furthermore, the ROI size on the tumor was defined at baseline and kept fixed for the analysis of data collected at following times. Thus, changes observed after therapy with respect to baseline may be due both to changes in tumor composition and tumor shrinkage.

### 3.1.2 Contralateral breast

Our results showed also a reduction in the concentration of total hemoglobin and in the oxygenation level (again both due to *HbO$_2$*) of the healthy contralateral breast by the end of therapy, when considering the average of CC and OB views (Fig. 4). Furthermore, in some subjects a decrease in water and collagen and an increase in lipid content was observed. This could be due to the systemic nature of NAC [48,49]. Indeed, the same was reported also by Altoe et al. [48], with a continuous wave diffuse optical imaging system, which allowed them to estimate blood parameters and water. In addition to DOS, this effect was confirmed using MRI by Chen et al. [49], who reported a reduction in breast density with NAC. Such reduction comes from a reduced amount of fibro-glandular tissue in favor of adipose tissue, which in turn corresponds to less water, collagen and total hemoglobin, and more lipids [12,50].

It is worth noting that these average variations are smaller compared to the ones in the tumor bearing breast, as shown in Table 3. In fact, except for oxy-hemoglobin (-46%, *p* = 0.0020), $SO_2$ (-20%, $p = 0.0020$) and scattering parameter $a$ (+11%, $p = 0.0098$), for other parameters an average change is observed, but there is no significant statistical difference between baseline and end-therapy. The marked reduction in the $HbO_2$ concentration might arise from the drugs used as part of the NAC regime, that can cause systemic damage of the lining of the blood vessels, leading to inflammation and decreased oxygen delivery to the tissues. These can also affect the production of red blood cells, which can determine anemia and decreased oxygen delivery to the tissue.

For the scattering parameters, the same consideration made for tumor tissue can be applied, but again for now full interpretation remains uncertain. Therefore, our primary focus lies on examining tissue absorption properties, as they ultimately provide insights into tissue chromophore concentrations.

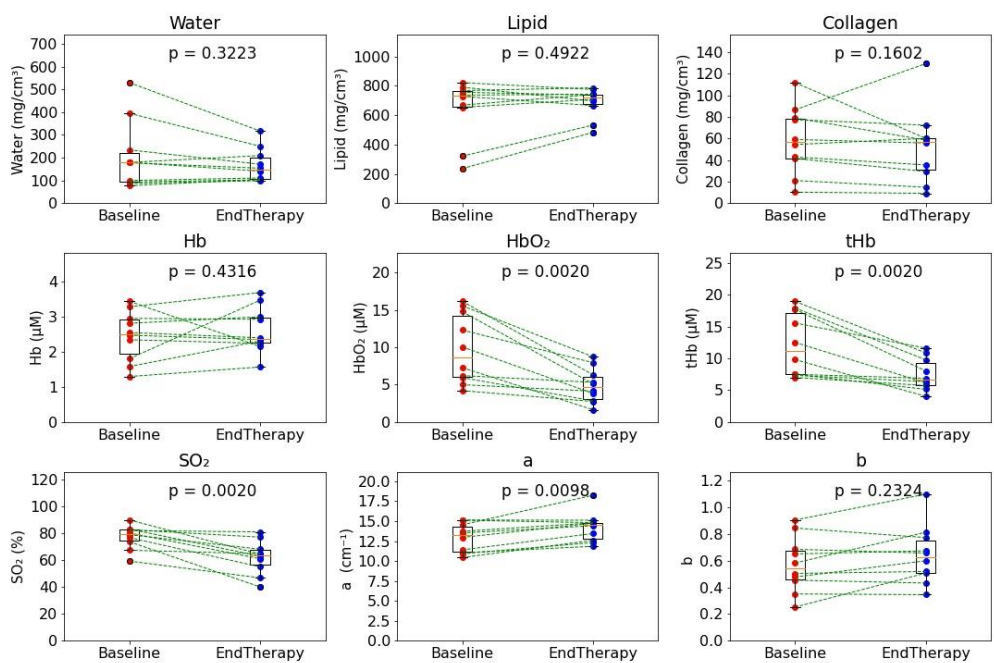

Fig. 4. Tissue constituents' concentrations, blood parameters, and scattering parameters at baseline and end of the therapy for 10 patients in contralateral healthy breast averaged over CC and OB views. *p*-values for the Wilcoxon signed-rank test are shown.

Table 3. Absolute tissue constituents' concentrations, blood parameters, and scattering parameters in contralateral healthy breast averaged across 10 subjects at baseline and end therapy sessions, and over CC and OB views. [% Change = [(EndTherapy_Concentration – Baseline_Concentration)/Baseline_Concentration].

|  | **Baseline** | **EndTherapy** | **Average change (%)** | ***p*-value** |
|---|---|---|---|---|
| **Water (mg/cm³)** | 205 ± 140 | 166 ± 69 | -5% | 0.3223 |
| **Lipid (mg/cm³)** | 648 ± 191 | 687 ± 96 | 16% | 0.4922 |
| **Collagen (mg/cm³)** | 58 ± 29 | 53 ± 33 | -11% | 0.1602 |
| **Hb (μM)** | 2.5 ± 0.7 | 2.6 ± 0.6 | 12% | 0.4316 |
| **HbO$_2$ (μM)** | 9.7 ± 4.4 | 4.9 ± 2.2 | -46% | 0.0020 |
| **tHb (μM)** | 12.2 ± 4.7 | 7.5 ± 2.4 | -35% | 0.0020 |
| **SO$_2$ (%)** | 77.5 ± 8.3 | 62.2 ± 11.8 | -20% | 0.0020 |
| *a* (cm$^{-1}$) | 12.87 ± 1.7 | 14.23 ± 1.8 | 11% | 0.0098 |
| *b* | 0.57 ± 0.20 | 0.64 ± 0.21 | 18% | 0.2324 |

*3.2     Potential role of Collagen in predicting response*

Few research groups have previously demonstrated oxy-hemoglobin's ability to predict chemotherapy response in a neoadjuvant setting [25,51–53]. However, collagen may present another intriguing avenue of exploration among various other constituents as our observations indicate that it experiences the most significant variation following oxy-hemoglobin by the conclusion of chemotherapy (Table 2).

Collagen is a critical component of the extracellular matrix (ECM), which is a complex network of proteins and carbohydrates that surrounds and supports cells in tissues. In breast cancer, collagen plays various roles in tumor development, progression, and metastasis [39,54]. During breast cancer development, there is significant remodeling of the ECM, including collagen [55]. Abnormal collagen deposition and organization can create an environment that promotes tumor growth and invasion. Collagen and the associated ECM components are being investigated as potential therapeutic targets. Some studies suggest that changes in collagen-related biomarkers in breast cancer patients could be used for prognostic purposes, predicting patient outcomes and treatment responses [56].

Despite being a crucial component of breast tissue, specific contribution of collagen has not been extensively explored or studied so far using the DOS method. The reason could be lack of suitable detection technology. Most studies are performed only below 900 nm, where more conventional and cheaper detectors are available. Detecting collagen itself might be difficult due to its low absorption in the near-infrared wavelength range and the overlap with the spectral characteristics of other more strongly absorbing tissue constituents. This makes it challenging to isolate collagen-specific signals. In this scenario, time domain diffuse optical spectroscopy offers an effective solution for assessing the optical properties of collagen because of its capability to naturally separate absorption from reduced scattering, allowing for the distortion-free evaluation of an absorption spectrum [57,58] and operation over a broad spectral range allow one to take advantage of spectral features of collagen, especially above 1000 nm, where cross-coupling with other absorbers is less marked..

Our research group at Politecnico di Milano has previously worked to estimate the collagen concentration in the breast tissue, to investigate its potential for lesion discrimination and breast density assessment [12,15,50,59]. These studies confirmed indeed the high potential of collagen for those applications. At the same time, it was also found that the inclusion of collagen in the fitting procedure has a significant impact on the estimated values of hemoglobin concentration, specifically deoxy-hemoglobin, due to a common trailing edge of their absorption spectra. This resulted in an average reduction in total hemoglobin content and higher values of oxygen saturation [57]. Such reduction in the estimated *Hb* content might make it more difficult to compare our results with those of other optical studies, which obtain a less complete assessment of tissue composition (e.g., excluding collagen or lipids). However, it is important to note that it remains consistent throughout the analysis, thus affecting only the absolute quantification and not the relative variations in this type of longitudinal studies or transversal comparisons.

Out of the 10 patients who completed their therapy, 4 were identified as complete responders, while 6 as partial responders based on their histopathological reports. Analysing their longitudinal data using the Student's *t*-test among 4 intervals, i.e., baseline, 2/3 weeks, mid-therapy and end therapy, it was observed that collagen is the only constituent in the tumor area that showed significant statistical difference (even if not a strong one) in complete responders ($p = 0.0423$) from baseline to 2/3 weeks and not in partial responders ($p = 0.1012$), as shown in Fig. 5. On the contrary, oxy-hemoglobin showed statistically significant decrease in both complete ($p = 0.0041$) and partial responders ($p = 0.0228$) already by the end of 2/3 weeks as shown in Fig. 6.

It is important to note that all patients underwent 6 measurement sessions, but dates may differ from the planned ones due to medical indisposition. In fact, sometimes the discomfort caused by drug toxicity prevented the patients from following the schedule strictly, especially the "2/5 days" measure. Consequently, the analysis presented in Fig. 5 is based on the available data points common to all the patients, specifically at the aforementioned 4 time intervals. The selected time points represent key stages of the treatment process and provide a comprehensive overview of the patients' responses to therapy. These results could be biased by the small number of patients and the absence of non-responders, but give an initial indication that collagen could be a biomarker to differentiate responders from partial responders, and supports interest in further investigation.

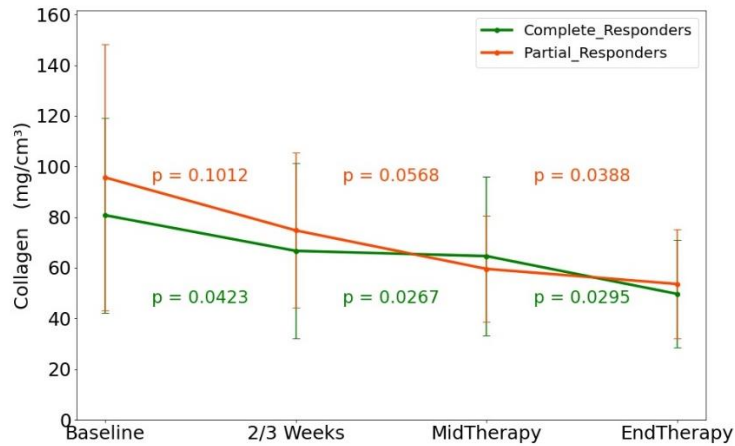

Fig. 5. Changes in average collagen concentration in complete responders (4 patients) and partial responders (6 patients) in green and orange lines, respectively. Data refer to tumor bearing breast averaged on CC and OB views at different points during course of NAC.

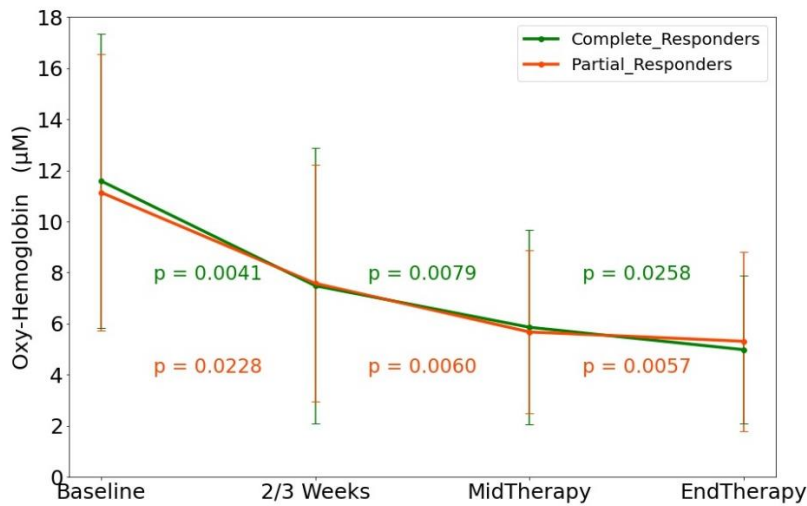

Fig. 6. Changes in average oxy-hemoglobin concentration in complete responders (4 patients) and partial responders (6 patients) in green and orange lines, respectively. Data refer to tumor bearing breast averaged on CC and OB views at different points during course of NAC.

**11**

## 4. Conclusion and Future Work

In summary, this is the first study to report changes in optically derived parameters measured by a time domain 7-wavelength optical mammograph to monitor the effect of NAC treatment on breast cancer with a more thorough estimate of tissue composition than reported up to now, at least to the authors' knowledge.

The changes in concentrations of total hemoglobin (-41% in tumor-bearing breast vs -35% in contralateral breast), water (-15% vs -5%) and collagen (-36% vs -11%) derived with the optical mammograph show decreasing trend on average over all patients (who responded at least partially to therapy), while lipids (+39% vs +16%) show increasing trend with therapy. These initial results demonstrate that our system exhibits sensitivity to changes in breast tissue composition induced by therapy, in line with the histopathological reports of our study cohort. The correlation between our quantitative measurements and histopathological evidence supports the potential clinical relevance and validity of our approach in assessing treatment response in breast cancer patients.

Some changes observed in the scattering properties cannot be fully explained with the change in tissue composition and suggest that NAC also causes some structural tissue changes that need to be further investigated.

In addition to the tumor breast, smaller but non-significant compositional changes except for the oxy-hemoglobin concentration were observed in the contralateral healthy breast, supporting systemic changes due to NAC.

In these initial results, it is observed that collagen is the only constituent that seems to distinguish complete responders from partial responders by the end of 2-3 weeks from the first NAC infusion, with a statistically significant variation with respect to baseline values ($p = 0.0423$ *vs* $p = 0.1012$). Therefore, we hypothesize that it could be a potential biomarker to measure NAC effectiveness.

The quantitative and qualitative behavior of changes in the constituents' concentrations in this type of neoadjuvant setting could provide physiological insight into patients' tumor and prognostication based on their response to therapy. DOS could provide non-invasively quantitative valuable information with regards to changes in tissue composition to the multidisciplinary team of healthcare professionals, helping them to predict response, tailor treatment options, personalize strategies and provide the best possible care for the patient in time.

The results presented here are preliminary in nature. In order to validate these findings, it is imperative to gather data from a more extensive pool of patients, in number and type (e.g., responders, partial responders, and non-responders; single and multi-regimen NAC therapy; neoadjuvant hormonotherapy and immunotherapy along with NAC; and histological type of tumor). The inclusion of a larger sample size will provide a more comprehensive and reliable assessment of the observed trends and help conjugate the overall average performance with the examination of specific cases.

We suppose that the optical data may also reveal regimen-dependent changes in breast tissue composition when a patient undergoes multi-drug and multi-cycle NAC treatment. With a larger cohort of patients available, we will be able to effectively analyze intermediate measurement sessions and investigate their correlation with the type of therapy.

Finally, currently our approach involves using a diffusion model for homogeneous media to interpret the data. However, we are aware of the inherent heterogeneity caused by tumor in breast tissue and are actively exploring the potential of analytical and numerical perturbative models, along with artificial intelligence approaches, to highlight contrasts, fully automatize the ROI selection and improve performances.

**Funding**

This work is being carried out under the PHAST project that has received funding from the European Union's Horizon 2020 research and innovation programme under grant agreement

No. 860185. The authors also acknowledge Fondazione Umberto Veronesi for partially funding the study.

**Disclosures**

The authors declare no conflicts of interest.

**Data availability**

Data underlying the results presented in this paper are not publicly available at this time but may be obtained from the authors upon reasonable request.